\documentclass[aps,pre,twocolumn,groupedaddress,showpacs]{revtex4}
\usepackage{graphicx}
\newcommand{\tri}{\triangle}

\newcommand{\cF}{{\cal F}}

\newcommand{\beq}{\begin{equation}}
\newcommand{\eeq}{\end{equation}}
\newcommand{\beqn}{\begin{eqnarray}}
\newcommand{\eeqn}{\end{eqnarray}}

\newcommand{\dd}{{\rm d}}
\newcommand{\ee}{{\rm e}}
\newcommand{\eq}{Eq.~}

\newcommand{\fig}{Fig.\ }

\begin{document}

\title{Self-assembly of protein amyloids:\\ a competition between amorphous 
and ordered aggregation
}
\author{Chiu Fan \surname{Lee}
}
\email{C.Lee1@physics.ox.ac.uk}
\affiliation{Physics Department, Clarendon Laboratory,
Oxford University, Parks Road, Oxford OX1 3PU, UK}


\date{\today}

\begin{abstract}
Protein aggregation in the form of amyloid fibrils has important biological and technological 
implications. Although the self-assembly process is highly efficient, aggregates not in the 
fibrillar form would also occur and it is important to include these disordered species when 
discussing the thermodynamic equilibrium behavior of the system.
Here, we initiate such a task by considering a mixture of monomeric proteins and the corresponding 
aggregates in the disordered form (micelles)
and in the fibrillar form (amyloid fibrils). Starting with a model on the respective binding free 
energies for these species, we calculate their concentrations at thermal equilibrium. We then 
discuss how the incorporation of the disordered structure furthers our understanding on the various 
amyloid promoting factors observed empirically, and on the kinetics of fibrilization.
\end{abstract}

\maketitle

\section{Introduction}
Amyloids are insoluble fibrous protein aggregations stabilized by a network of hydrogen bonds and 
hydrophobic interactions \cite{Sunde_JMB97, Dobson_Nature03, 
Radford_TrendsBiochemSci00,Sawaya_Nature07}.
They are intimately related to many neurodegenerative diseases such as the Alzheimer's Disease, the 
Parkinson Disease and other prion diseases \cite{Harper_AnnuRevBiochem97}.
Better characterization of the various properties of amyloid fibrils is therefore of high 
importance for the understanding of the associated pathogenesis.
More recently, viewing protein amyloid formation as a highly efficient self-assembly process, 
possible applications have also been proposed. For instance,
amyloid fibrils have been employed as nanowire templates \cite{Reches_Science03,Scheibel_PNAS03}, 
were shown to possess great tensile strength \cite{Knowles_Science07,Smith_PNAS06} and complex 
phase behavior similar to liquid crystals \cite{Corrigan_JACS06, Lee_a09b}.   Given the high 
importance of protein amyloid in biology and potentially in technology, it is being studied 
intensively. In particular, much effort has been spent on investigating the amino-acid dependency 
on amyloid propensity \cite{Reches_JBiolChem02, Reches_Science03,Tracz_Biochem04,  
Ma_CurrOpinChemBiol06,Bemporad_ProtSci06,Marek_Biochem07,  Jean_PLoSONE08};
the  possibility of primary-sequence-based amyloid propensity predictions \cite{Yoon_ProtSci04,
Fernandez_NatBiotech04,Tartaglia_ProtSci05,Galzitskaya_PLoSCompBiol06,DuBay_JMB04,Pawar_JMB05}; the 
mechanical properties of protein amyloid \cite{Smith_PNAS06, Knowles_PRL06,Knowles_Nanotech07}; the 
kinetics of amyloid formation 
\cite{Serio_Science00,Yong_PNAS02,Nguyen_PNAS04,Pellarin_JMB06,Nguyen_PNAS07,Pellarin_JMB07,Cheon_P
LoSCompBiol07,Xue_PNAS08,Zhang_JChemPhys09}; as well as the thermodynamical behaviours of the 
aggregation process \cite{vanGestel_BiophysJ06, Aggeli_PNAS01,Nyrkova_EPJB00, Tiana_JChemPhys04}.

\begin{figure}
\caption{(Color online)
Schematic diagrams of the three species considered in this paper: (a) a monomeric protein in 
solution; (b) a micelle, or an amorphous aggregate; (c) an eight-monomer segment of an amyloid 
fibril consisting of two cross-beta structures (one cross-beta structure is coloured, the other is 
black). The hydrogen bonds stabilise the beta sheets in the vertical direction (not shown in this 
figure). (Drawn with DeepView 
\cite{Guex_Electophoresis97}.)}
\label{Pic}
\begin{center}
\includegraphics[scale=.5]{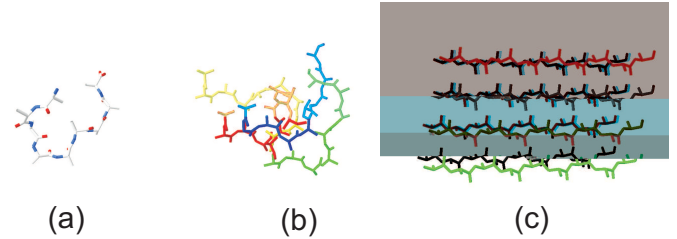}
\end{center}
\end{figure}

Although the protein amyloid self-assembly process is highly efficient, aggregates not in the 
fibrillar form would also occur and it is important to include these disordered species when 
discussing the thermodynamic equilibrium behavior of the system. This motivates us to consider here 
a system consisting of a mixture of monomers, aggregates with a linearly ordered structure 
(fibrils) and aggregates with a disordered structure (a micelle-like aggregate) (c.f.\ \fig 
\ref{Pic}). Starting with a discussion on their respective binding free energies, we deduce the 
concentrations for the various species at thermal equilibrium, and consider the experimental 
implications of our investigation. In particular, we study the effect of temperature and pressure 
variations on the average fibrillar length. We then discuss how our work relates to the empirically 
observed variation in amyloid propensity with
respect to the primary sequences of the proteins. Finally, we 
employ the formalism developed to study the kinetic process of aggregation.

The plan of the paper is as follows: 
In Section II, we introduce our model of a amyloid-forming self-assembly system.  In Section III, 
we discuss the experimentally relevant predictions from our model. In Section IV, we consider how 
our findings relate to empirical observations on amyloid propensity. In Section V, we investigate 
the kinetic process of self-assembly from the perspective of our free energy picture.

\section{The model}
\label{model}
In this work, we are primarily concerned with amyloid fibrilization of short peptides. Peptides 
interact via an array of interactions, such as hydrophobic interactions, hydrogen bonding, 
electrostatic interactions, etc (for a review, see, e.g., 
\cite{Sneppen_B05,Jackson_B06}).  Due to these interactions, aggregation may occur and we  consider 
here two different types of aggregates: i) linearly structured aggregates (amyloid fibrils) and ii)
disordered aggregates (micelles) (c.f. Fig.~\ref{Pic}). For the micellar
species, we assume that there is an optimal configuration consisting of $M$ 
proteins, where $M$ is in the order of tens \footnote{For instance, $M$ is in the order of 30 for 
amyloid-$\beta$ as demonstrated exprimentally in \cite{Yong_PNAS02}.}. 
For the fibrillar species, we assume that the only ordered structure is a two-tape structure, i.e., 
each fibril consists of stacking two cross-beta structures (c.f. Fig.~\ref{Pic}(c)). We note that 
amyloid fibrils can exhibit structural variations even when prepared under the same condition, and 
the precise structural details will be highly primary sequence dependent (see, e.g., 
\cite{Jimenez_PNAS02}).

Now a note on terminology: we will call a free protein in solution a monomer, 
and a fibril consisting of $i$ proteins a $i$-mer fibril. We will also denote from now on the 
numbers of monomers, micelles and $i$-mer fibril in a solution of volume $V$ by $N^{(a)}$, 
$N^{(b)}$ and $N^{(c)}_i$ respectively. In particular, if $N$ denotes the total number of monomers, 
we have
\beq
\label{constraint}
N^{(a)}+M N^{(b)}+\sum_i i N^{(c)}_i = N \ .
\eeq

Given the three different species: monomers, micelles and fibrils, we are interested in determining 
their respective volume fractions given an initial volume fraction $C$. In this work, we will 
always set the unit of volume to be the volume of one monomer. With this convention, $C = N/V$ 
where $V$ is the volume of the system.

To calculate the relative abundances for the various species, we firstly need to obtain their 
respective species-specific Binding Free Energy (BFE). Without loss of generality, we will set the 
monomeric BFE to zero, and denote the micellar BFE by $\gamma$, where
\beq
\gamma =-T \tri s_b +\tri \epsilon_b +p\tri v_b \ .
\eeq
In the above equation, $\tri s_a$, $\tri \epsilon_b$ and $\tri v_b$ are the entropic, binding 
energy and volumic differences between the monomers and the micelles. In other words, $\tri s_b$ 
quantifies the free energy contribution from the loss in configurational freedom due to the 
rigidity of the aggregates, $\tri \epsilon_b$ quantifies the change in internal energy resulting 
from the various inter-protein and intra-protein interactions, and $\tri v_b$ denotes the change in 
volume for a monomer as a result of being part of a larger aggregate.

For the fibrillar species, we will denote the BFE of an infinitely long fibril as:
\beq
\label{finf}
f_\infty = -T \tri s_c +\tri \epsilon_c +p\tri v_c \ .
\eeq
The terms in the R.H.S.\ above have similar definitions as in the micellar case aforementioned. For 
a finite-size $i$-mer fibril, we will make the following assumption typically made in the study of 
linearly aggregating systems (e.g., see \cite{Israelachvili_B91}):
\beq
f_i = \left(\frac{i-\xi}{i}\right) f_\infty \ ,
\eeq
where $\xi$ $(\xi >0)$ accounts for the boundary effect at the fibril's ends and is of order one
\cite{Israelachvili_B91}. For instance, it accounts for the loss of hydrophobic interactions at 
both ends of a fibril.

With the BFE defined, we can calculate the concentrations of the various species by finding the 
minimum of the total free energy of the system. We will start by writing the total partition 
function as \cite{Mutaftschiev_B01}:
\beq
\label{Qtot}
Q_{\rm tot} = \prod'_{i} \frac{(AV)^{ N^{(a)}}(B V)^{N^{(b)}}(C_i V)^{N^{(c)}_i}}{N^{(a)}!\
N^{(b)}!\ N_i^{(c)}!}
\eeq
where, by the previous discussion on the BFE \footnote{Note that 
$A$, $B$ and  $C_i$ in \eq (\ref{Qtot}) correspond to the partition functions for a monomer, a 
micelle and a $i$-mer fibril respectively. For instance,
$
B = \int_{\Lambda_b} \dd [q] \ee^{ -U(\{q\})/k_BT}
$
where $\Lambda_b$ denotes the volume of the coordinate space defining a micelle, the set $\{ q \}$ 
denotes the set of degrees of freedom of the aggregates, i.e., the coordinates of all the atoms 
making up the micelle, and $U(\{q\})$ is the internal energy of the system.
},
\beqn
A &=& 1
\\
B &=& \exp(-M\gamma/k_BT)
\\
C_j &=& \exp[-(i-\xi)f_\infty/k_BT] \ .
\eeqn
Note that the prime in the product in \eq (\ref{Qtot}) denotes the restriction that the total 
number of peptides is conserved (c.f.~\eq (\ref{constraint})).
The distribution of the various species can now be obtained by determining the minimum of the total 
free energy density, 
\beq
\cF_{\rm tot} = -\frac{k_BT \log Q_{\rm tot}}{V} \ ,
\eeq
subject to the constraint shown in \eq (\ref{constraint}). This optimization problem can be solved 
by the Lagrange multiplier method and the results are (see, e.g., \cite{Schoot_Langmuir94}):
\beqn
\label{CMC}
n^{(b)} &=& \left[n^{(a)}\ee^{-\gamma /k_BT} \right]^M
\\
\label{CFC}
n^{(c)}_i &=& \left[n^{(a)}\ee^{-(i-\xi)f_\infty/i k_BT} \right]^i \ .
\eeqn
The lower case $n$ denotes  the volume fraction of the corresponding species, i.e., $n^{(.)} \equiv 
N^{(.)}/V$.

For the micellar species, due to the magnitude of $M$ ($M \geq 10$ [56]), if 
$C< \exp(\gamma /k_BT)$, the micellar volume fraction will be negligible in comparison to the 
monomer volume fraction; conversely, if $C>\exp(\gamma /k_BT)$, then the monomeric volume fraction 
will be $\exp(\gamma /k_BT)$ and all the excess monomers will be in the micellar form, i.e.,
$n^{(b)} \simeq C-\exp(\gamma /k_BT)$  \cite{Israelachvili_B91}. It is therefore legitimate to 
define a critical concentration at $C_{\rm crit}=\exp(\gamma /k_BT)$. We will call this the 
Critical Micellar Concentration (CMC). For the fibrillar species, a similar reasoning indicates 
that the critical concentration for fibrilization is at $C_{\rm crit} = \exp(f_\infty/k_BT)$. We 
will call  this the Critical Fibrillar Concentration (CFC). 

\begin{figure}
\caption{(Color online)
Dominance diagram of the three-species system at concentration higher than the critical 
concentrations: CMC and CFC.   The coloured arrows depict how the dominance may shift under 
increase in hydrophobicity (red), increase in the number of aromatic side chains (blue), increase 
in alternating hydrophobic-hydrophilic amino-acid sequence (green), and increase in unpaired 
charges in the side chains (black).
{\it Inset plot:} The volume fractions of $i$-mer fibrils versus $i$ in the fibrillar phase (i.e., 
$C \gg {\rm CFC}$ and  ${\rm CMC} > {\rm CFC}$), where $L \equiv \langle iN^{(c)}_i \rangle$ is set 
to be 500.
}
\label{phase}
\begin{center}
\includegraphics[scale=.45]{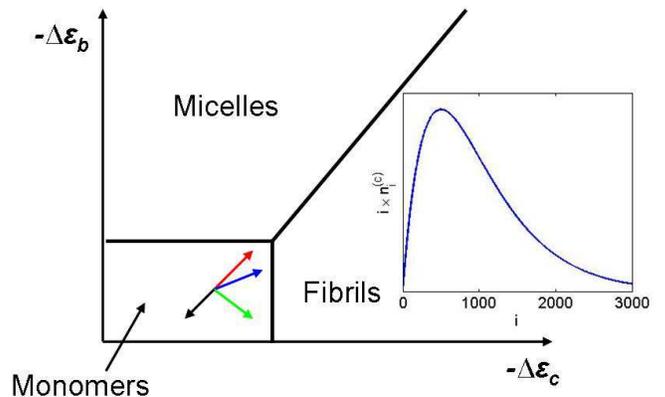}
\end{center}
\end{figure}

If CMC $<$ CFC and $C \gg$ CMC, CFC,  almost all monomers would be in the micellar form and the 
concentrations of monomers and fibrils are negligible by comparison. On the other hand, if $C \gg 
{\rm CFC}, {\rm CMC}$ and CFC $<$ CMC, then the concentrations of monomers and micelles will be 
negligible while the concentration of the fibrillar species 
will be abundant. In this fibril-dominant regime,  $n^{(c)}_i$ follows  the following distribution 
\cite{Israelachvili_B91}:
\beq
\label{distribution}
n^{(c)}_i = \exp[ -i /L + \xi f_\infty/k_BT]
\eeq
where $L$ is the average number of monomers in a fibril such that
\beq
\label{aveL}
L \equiv \langle i N^{(c)}_i \rangle = \sqrt{C \ee^{-\xi f_\infty/k_BT}} \ .
\eeq
Since a fibril is a linear structure, the average fibrillar length is thus proportional to $L$. 
According to \eq (\ref{aveL}), the average fibril length scales with $\sqrt{C}$. This fact is 
observed in other linear aggregating systems  and is a manifestation of the one-dimensional nature 
of the aggregates \cite{Israelachvili_JFaraday76, Israelachvili_B91}. 
The profile of $i\times n^{(c)}_i$ versus $i$ is depicted in the inset plot in \fig \ref{phase}. 
This analytical result is qualitatively confirmed by the experimental
observations on $\beta$-lactoglobulin amyloid fibrils
\cite{Rogers_Macromol05,Rogers_EPJE05}.
\vspace{.2in}

We will now try to estimate the magnitudes of the terms appearing in \eq (\ref{finf}).  For the 
first term,
let us assume that a protein in monomeric form is in the denatured state, and a fibrilized protein 
corresponds to the folded state. It has been experimentally and theoretically estimated that,
by going from the denatured state to the folded state,
 a protein loses on average around $k_B \ln 10$ per amino acid in entropy 
\cite{Brandts_JACS64,Bryngelson_PNAS87,Sneppen_B05}. We will therefore estimate $\tri s_c$ as 
$-Rk_B \ln 10 = -2.3\times R k_B$ where $R$ is the number of amino acids in the protein . 
For the third term in \eq (\ref{finf}), it has been demonstrated that the change in the protein's 
volume upon folding is very small \cite{Harpaz_Struct94}. Indeed, it is found  that the change in 
volume per amino acid upon folding is in the order of 0.01 nm$^3$ \cite{Harpaz_Struct94}, which 
suggests that $p\tri v_c \sim 0.07\times Rk_BT$ at atmospheric pressure. It is therefore negligible 
in comparison to the entropic contribution. The second term in \eq(\ref{finf}) involves a 
combination of interactions, such as hydrogen bonding, hydrophobic interactions, electrostatic 
interactions, etc, among which hydrophobic interactions, 
which are of the order of a few $k_BT$ per amino acid,
 are believed to be dominant \cite{Dill_Biochem90, Jackson_B06}. 
Since hydrophobic interactions involves effective burying of hydrophobic side-chains inside the 
protein structure, it indicates the need for a multi-layered fibrillar structure (such as our 
two-tape model employed here), as universally observed in amyloid fibrils formed from different 
proteins \cite{Sawaya_Nature07}.

\section{Experimental implications}
We will now focus on the fibrillar phase, i.e., we are in the scenario where $C > $ CFC  and CFC 
$<$ CMC. According to \eq (\ref{aveL}):
\beq
\label{aveL2}
\ln L = -\frac{\xi f_\infty}{2k_B T}  + \frac{1}{2} \ln C \ .
\eeq
If we equate a monomeric 
protein to a denatured protein, and a fibrilized protein to a folded protein, then experimental 
work indicates that $f_\infty/T$ is a concave up function with respect to $T$ such that the minimum 
occurs at around 20$^\circ$C \cite{Schellman_BiophysJ97, Baldwin_PNAS86}. This suggests that in an 
isobaric experiment, the average fibril length would first increase and then decrease as 
temperature increases.

The situation for pressure variation is more complicated due to the fact that the compressibility 
differs for different amino acids. Nevertheless, it has been found generally  that at low pressure 
($\sim$ 1 atm), the change in volume upon folding is small while the change is positive at very 
high pressure (7500 atm \cite{Hawley_Biochem71}) due to the fact that denatured protein has greater 
compressibility \cite{Harpaz_Struct94, Royer_BBA02}. In other words, if we again equate a monomeric 
protein to a denatured protein, and a fibrilized protein to a folded protein, we would expect that, 
in the very high pressure regime, an increase in pressure would lead to an 
exponential decrease in the average fibrillar length 
in an isothermal experiments.

\section{Relevance to previous empirical findings}
As discussed in Sect.\ \ref{model}, if $C \gg {\rm CMC, CFC}$, the dominant species in the system 
will be the one with a lower critical concentration. Namely,
in terms of BFE,  the dominant species will be fibrillar if $f_\infty < \gamma$, and vice versa. We 
will now discuss how the primary sequence may affect amyloid propensity in terms of the BFE. To 
simplify the discussion, we will assume that substituting an amino acid affects predominantly the 
binding energy term, $\tri \epsilon$, in the BFE (c.f.\ \fig \ref{phase}). 

As a result of empirical observations \cite{Yoon_ProtSci04,
Fernandez_NatBiotech04,Tartaglia_ProtSci05,Galzitskaya_PLoSCompBiol06,DuBay_JMB04,Pawar_JMB05}, it 
is generally agreed that the following factors promote amyloid formation:
i) an increase in hydrophobicity, and ii) an increase in length of an alternating 
hydrophobic-hydrophilic amino-acid sequence; while it is found that an increase in the number of 
charged amino acids decreases amyloid propensity. In terms of the binding energies, an increase in 
hydrophobicity would decrease both $\tri \epsilon_b$ and $\tri \epsilon_c$ and as such would on 
average increase the fibrilization probability if the protein's parameters are already close to the 
monomer-fibril boundary in the dominance diagram (the red arrow in the Fig.~\ref{phase}). For our 
two-tape model for the amyloid fibril (c.f. Fig.~\ref{Pic}(c)), an increase in alternating 
hydrophobic-hydrophilic amino-acid sequence would allow for packing the hydrophobic side chains 
inside the cross-beta sheet structure, while having the hydrophilic side chains outside, this would 
decrease $\tri \epsilon_c$. On the other hand, having such a pattern would conceivably decrease the 
average energy gained inside a micellar structure given the amorphous structural nature, i.e., 
$\tri \epsilon_b$ will be increased. Such a modification would therefore increase amyloid 
propensity (the green arrow in the Fig.~\ref{phase}). If there is an increase in paired charges in 
the protein, i.e., charges that are not accompanied by ionic bonds, electrostatic interaction would 
deter aggregation and as such both $\tri \epsilon_b$ and $\tri \epsilon_c$ will be increased (the 
black arrow in the Fig.~\ref{phase}).

\begin{figure}
\caption{(Color online)
(a) A schematic digram depicting the amyloid-beta self-assembly process proposed in 
\cite{Lomakin_PNAS97}. The circle denotes the free monomeric state, the square denotes the typical 
micellar state (the micellar size, $M$, is estimated to be 25 \cite{Lomakin_PNAS97}), and the 
triangle denotes a stable nucleus (the nucleus size is estimated to be 10 \cite{Lomakin_PNAS97}). 
The thick arrow depicts the fast pathway from free monomer to micelles and the thin arrow depicts 
the slow process of nucleation from micelles. The broken arrow depicts that very slow process of 
nucleation from free monomers, which is out of the range of experimental time scale, but may play 
an important role in actual pathogenesis under physiological time scale.  
(b) The temporal evolution of monomer concentration. {\it Upper plot:} When $C > {\rm CMC}$, the 
monomers are quickly converted into micelles and then slowly into fibrils. The figures above the 
curve depict the dominant species in the solution as time progresses.
{\it Lower plot:} When ${\rm CFC} < C< {\rm CMC}$, the proteins remain in monomeric form for a time 
longer than can be probed experimentally. These two plots show the curious phenomenon of the 
possibility of ending up with a lower monomeric concentration when the initial concentration is 
higher.
}
\label{dynamics}
\begin{center}
\includegraphics[scale=.45]{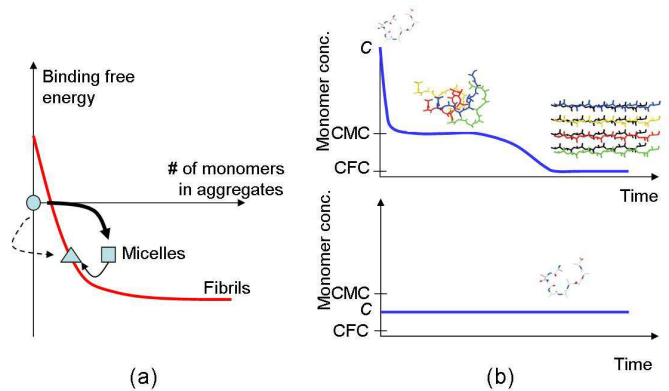}
\end{center}
\end{figure}

Another insight we can gain from the above consideration concerns the importance of aromatic 
residues in amyloid propensity \cite{Reches_JBiolChem02, Reches_Science03,Tracz_Biochem04,  
Ma_CurrOpinChemBiol06,Bemporad_ProtSci06,Marek_Biochem07,  Jean_PLoSONE08}. Beside the heightened 
hydrophobicity in aromatic residues, the offsetted $\pi$-stacking interaction is directional along 
the fibrillar axis \cite{Hunter_JMB91,Meyer_AngeChem03}, hence $\epsilon_c$ may be decreased more 
than $\epsilon_b$ (the blue arrow in the Fig.~\ref{phase}). This suggests that aromatic 
interaction, or any interactions directional along the fibrillar axis, contributes to amyloid
stability in a way different from ordinary hydrophobic interactions.

\section{Kinetics}
We  discuss now how the picture developed in this paper helps to describe the kinetics of the 
protein amyloid self-assembly process investigated experimentally. According to the model proposed 
in \cite{Lomakin_PNAS97}, the series of events leading up to the fibrilization of amyloid-$\beta$ 
proteins is depicted in \fig \ref{dynamics}. In this scenario, the direct pathway from monomers to 
stable nucleus (depicted by the broken arrow in \fig \ref{dynamics}) is in a time scale too long to 
be probed experimentally. Therefore, the only possible fibrilization pathway is for the monomers to 
first formed micelles (a fast process, depicted by the thick black arrow), stable nuclei are then 
formed out of the micelles (a slow process, depicted by the thin black arrow). 
Based on this model, within the temporal constraint of experiments, fibrilization is only possible  
if $C > {\rm CMC}$ (c.f.\ \fig \ref{dynamics}(b)). In the case of the amyloid-$\beta$ protein, the 
CMC has been measured to be in the order of 10 $\mu$M \cite{Terzi_Biochem97}. This is substantially 
higher than the concentration of amyloid-beta in the cerebral spinal fluid, which is in the 
subnanomolar concentration range \cite{Seubert_Nature92}. It therefore poses the question are 
current experimental methods only probing the fast pathway -- monomers to micelles to nucleus 
(depicted by the two solid arrows), while the physiologically relevant pathway is the slow pathway 
-- monomers to nucleus (depicted by the broken arrow).

\section{Conclusion}
In this work, we have considered the thermodynamic equilibrium behavior of a system with a mixture 
of monomeric proteins, the corresponding micellar aggregates and fibrillar aggregates. We have 
deduced the concentrations of these species at thermal equilibrium and we have found that the 
average fibrillar length is very sensitive to temperature or pressure variation.
We have also discussed the relevance of our investigation to previous empirical findings and to the 
understanding of the kinetical  process of fibrilization.

\begin{acknowledgements}
The author thanks Catherine Davison, L\'{e}titia Jean and David Vaux at the Dunn School of 
Pathology (Oxford) for many stimulating discussions, and the Glasstone Trust (Oxford) and Jesus 
College (Oxford) for financial support.
\end{acknowledgements}


\end{document}